\documentclass[runningheads]{llncs}
\usepackage[T1]{fontenc}
\usepackage{amsmath,amssymb}
\usepackage{graphicx}
\usepackage{booktabs}
\usepackage{microtype}

\usepackage[hidelinks]{hyperref}
\usepackage{cleveref}
\graphicspath{{figures/}}

\usepackage{xcolor}

\begin{document}
\title{Stable Self-Modulating Quantum Fast-Weight Programmers with Bounded Memory Gates}
\titlerunning{Stable Self-Modulating QFWPs with Bounded Memory Gates}
\author{Kuo-Chung Peng\inst{1}\orcidID{0009-0001-8342-2481}
\and
Jiun-Cheng Jiang\inst{1}\orcidID{0009-0005-1134-4962}
\and
Chun-Hua Lin\inst{1}\orcidID{0009-0002-4383-0453}
\and 
Yifeng Peng\inst{2}\orcidID{0009-0007-3306-9417}
\and
Junghoon Justin Park\inst{3}\orcidID{0000-0001-8982-0387}
\and 
Huan-Hsin Tseng \inst{4}\orcidID{0000-0001-9544-4226}
\and 
Hsin-Yi Lin \inst{4}\orcidID{0000-0001-5731-2353}
\and 
Kuan-Cheng Chen \inst{5}\orcidID{0000-0002-6575-7034}
\and
Chen-Yu Liu \inst{1}\orcidID{0000-0002-5437-5188}
\and
Shinjae Yoo \inst{4}\orcidID{0000-0003-4378-6448}
\and 
Samuel Yen-Chi Chen \inst{6}\orcidID{0000-0003-0114-4826}}
\authorrunning{K.-C. Peng et al.}
\institute{
National Taiwan University, Taiwan 
\and
Stevens Institute of Technology, NJ, USA
\and
Seoul National University, Korea
\and
Brookhaven National Laboratory, NY, USA
\and
Imperial College London, UK
\and
Wells Fargo, NY, USA
\\
\email{ycchen1989@ieee.org}}
\maketitle              %
\begin{abstract}
Quantum Fast-Weight Programmers (QFWPs) store temporal information in dynamically programmed variational-circuit parameters rather than in nonlinear recurrent hidden states, offering a practical route to quantum sequence modeling. Self-Modulating QFWP improves this framework by using input-dependent gates for both new fast-weight updates and the accumulated fast-weight state, but its unbounded old-state multiplier can diverge in long-sequence regimes. We propose a bounded old-state modulation rule that applies a sign-preserving \(\tanh\) gate only to the recurrent memory branch while leaving the additive update and new-update modulation unchanged. We evaluate standard QFWP, full Self-Modulating QFWP, Only-New, and Only-Old variants on two CUDA-Q quantum-dynamics forecasting tasks and on Milan SMS telecommunication activity prediction. The quantum-dynamics results show that old-state modulation is the most consistent source of improvement over Standard QFWP, and that bounding the old-state gate removes long-sequence divergence while improving aggregate robustness. On Milan SMS forecasting, the original unbounded Self-Modulating QFWP converges across the tested grid and shows its clearest gains at longer input windows, with behavior close to the Only-Old ablation. These findings identify accumulated-memory modulation as the key mechanism of Self-Modulating QFWP and bounded old-state gating as a targeted stabilization strategy.

\keywords{Quantum Fast-Weight Programmer \and Self-Modulating QFWP \and
Bounded Memory Gate \and Quantum Time-Series Forecasting}
\end{abstract}

\section{Introduction}
\label{sec:introduction}

Quantum sequence models have become a promising direction for learning temporal dependencies with hybrid quantum--classical architectures\cite{chen2025benchmarking,anschuetz2023interpretable,bausch2020recurrent}. A representative starting point is the Quantum Long Short-Term Memory (QLSTM) model, which replaces parts of the classical LSTM cell with variational quantum circuits and has shown promising empirical behavior on temporal learning tasks~\cite{chen2022qlstm,chen2025toward,zhang2026quantum,chen2026recursive_qlstm}. These studies suggest that variational quantum circuits can act as compact nonlinear processors for temporal data, but recurrent quantum models still inherit a central limitation during training. Models with nonlinear recurrence must propagate gradients through a time-ordered nonlinear state evolution.

Quantum Fast Weight Programmers (QFWPs) were introduced to address this bottleneck by moving memory from a recurrent hidden state into dynamically programmed circuit parameters~\cite{chen2024qfwp}. In QFWP, a classical slow programmer reads each input and generates an update to the parameters of a variational quantum circuit, which acts as the fast programmer. Because the temporal state is represented by accumulated fast weights rather than by a nonlinear recurrent hidden state, the model avoids backpropagation through time (BPTT) across a quantum recurrent cell and admits a simpler, more parallelizable gradient path~\cite{peng2026qkanfwp}. This makes QFWP an attractive framework for quantum time-series prediction and sequential control, where circuit-evaluation cost and gradient depth are major practical constraints.

The recent Self-Modulating QFWP extends this framework by introducing input-dependent multiplicative modulation over both the newly generated fast-weight update and the previously accumulated fast-weight state~\cite{chen2026selfmod}. This update rule gives the model direct control over how new information is injected and how past fast-weight memory is retained, suppressed, amplified, or sign-reversed. Prior results indicate that such self-modulation improves convergence and prediction accuracy, with the old-state branch often providing the dominant source of improvement~\cite{chen2026selfmod}. However, because the original old-state multiplier is unconstrained, repeated multiplicative updates can become unstable in long-sequence regimes.
In this work, we revisit Self-Modulating QFWP on two complementary next-step prediction settings: telecommunication traffic prediction and quantum-dynamics prediction. The former tests the model on practical temporal signals, while the latter evaluates whether adaptive fast-weight memory can learn observables generated by simulated quantum systems. We compare Standard QFWP, full Self-Modulating QFWP, and the Only-New and Only-Old Self-Modulating ablations across hidden sizes and input-window lengths, and we use relative-improvement, old-state strength, and synergy diagnostics to identify which modulation branch drives the observed gains.

Our main architectural contribution is a bounded old-state modulation rule. Specifically, we apply a sign-preserving \(\tanh\) bound to the old-state multiplicative gate while leaving the additive update and new-update modulation unchanged. This isolates the stabilization mechanism to the recurrent fast-weight memory branch. Empirically, the bounded gate preserves the low-error behavior of old-state modulation while curing the long-sequence divergence observed in unbounded multiplicative variants. The resulting analysis supports a focused conclusion: input-dependent control of accumulated fast weights is central to the benefit of Self-Modulating QFWP, and bounding this control provides a simple and effective route to stable quantum fast-weight programming.

\section{Related Work}
\label{sec:related_work}

\subsection{Quantum and quantum-inspired sequential models}
Early QRNN and QLSTM studies showed that 
variational quantum circuits (VQCs) can process temporal data and sequence-recognition tasks~\cite{bausch2020recurrent,chen2022qlstm}.
Recent works also investigate VQC-instantiated quantum-inspired Kolmogorov--Arnold network (QKAN)-based LSTM models and transformer models implemented by single-qubit data re-uploading circuits for scalable yet efficient quantum-inspired sequence learning \cite{jiang2025qvaf_qkan,hsu2026qkanlstm,lin2026gqkae}.
Applications have since expanded to environmental and energy forecasting, including climate~\cite{hsu2025quantum}, air-quality~\cite{li2024airqlstm}, solar-power~\cite{khan2024solarqlstm}, and flood prediction~\cite{lin2024qtlstm}; economic and infrastructure forecasting, including carbon prices~\cite{cao2023linear}, stock indices~\cite{su2025blsqlstm}, and urban telecommunication traffic~\cite{chen2025benchmarking,hsu2026qkanlstm}; and scientific or biomedical sequence modeling, including drug discovery~\cite{zhang2026quantum}, predictive maintenance~\cite{tsurkan2025hqrnn}, human activity recognition~\cite{hsu2025federated_qkernel_lstm}, and wearable-health estimation~\cite{tran2025eqlstml}. These studies demonstrate the breadth of quantum sequential modeling, but recurrent quantum cells still couple computation across time and therefore retain the cost of sequential backpropagation through time. 

\vspace{-5pt}
\subsection{Fast weight programmers}

Fast Weight Programmers (FWPs) take a different view of memory: a slow network writes context-dependent parameters into a fast network, so temporal information is stored in fast-weight dynamics rather than in a nonlinear hidden-state recurrence~\cite{schmidhuber1992fast}. This classical idea was later connected to linear self-attention and extended through recurrent fast-weight variants~\cite{schlag2021linear,irie2021going}. QFWP transfers the FWP principle to hybrid quantum learning by using a classical slow programmer to update the parameters of a variational quantum circuit, with demonstrations on time-series prediction and reinforcement learning~\cite{chen2024qfwp,ceschini2026qfwp}. Recent extensions include Quantum-Train QFWP for parameter-efficient circuit programming~\cite{liu2025qtqfwp} and quantum-inspired Gated QKAN-FWP for scalable sequence learning, long-horizon solar-cycle prediction, MiniGrid reinforcement learning~\cite{peng2026qkanfwp}, and traffic matrix forecasting~\cite{peng2026qkanfwp_TM}. The prior Self-Modulating QFWP further introduces input-dependent gates on both the new fast-weight update and the accumulated fast-weight state~\cite{chen2026selfmod}; our work preserves this memory-control idea but bounds the old-state multiplier to remove long-sequence divergence.

\section{Model}
\label{sec:model}

\subsection{QFWP baseline}
\label{subsec:qfwp}

Let \(Q\) denote the number of qubits and \(L\) the number of trainable variational layers in the fast variational quantum circuit. At time step \(t\), the fast circuit parameters form a matrix \(\Theta_t\in\mathbb{R}^{L\times Q}\). A classical slow controller maps the scalar input \(x_t\) to a hidden state
\begin{equation}
  h_t = \phi_{\Omega}(x_t)\in\mathbb{R}^{H},
\end{equation}
where \(\Omega\) are trainable controller parameters. Two affine heads generate vectors
\begin{equation}
  \ell_t = W_\ell h_t + b_\ell \in \mathbb{R}^{L},
  \qquad
  r_t = W_r h_t + b_r \in \mathbb{R}^{Q},
\end{equation}
and their outer product gives the raw fast-weight update
\begin{equation}
  \Delta_t = \ell_t r_t^{\top} \in \mathbb{R}^{L\times Q}.
  \label{eq:delta}
\end{equation}
The additive QFWP baseline is
\begin{equation}
  \Theta_t = \Theta_{t-1}+\Delta_t.
  \label{eq:standard}
\end{equation}
The variational circuit then uses \(\Theta_t\) to produce quantum expectation-value features for the next-step prediction head. In all experiments below, the controller width and the number of qubits are tied, \(H=Q\), and the circuit depth is fixed at \(L=5\) for quantum dynamics tasks and \(L=2\) for the telecommunication task. For the full circuit construction and implementation details, we follow prior works~\cite{chen2024qfwp,chen2026selfmod}.

\subsection{Self-modulation inherited from the previous model}
\label{subsec:selfmod}

Self-Modulating QFWP adds two input-dependent modulation matrices, one for the new update and one for the old accumulated state. For \(s\in\{\mathrm{new},\mathrm{old}\}\), the modulation head produces
\begin{equation}
  m_t^{s,L}=W^{s,L}h_t+b^{s,L}\in\mathbb{R}^{L},
  \qquad
  m_t^{s,Q}=W^{s,Q}h_t+b^{s,Q}\in\mathbb{R}^{Q},
\end{equation}
and
\begin{equation}
  M_t^s = m_t^{s,L}(m_t^{s,Q})^{\top}\in\mathbb{R}^{L\times Q}.
  \label{eq:gate}
\end{equation}
The unbounded full self-modulating update is
\begin{equation}
  \Theta_t = \Delta_t\odot M_t^{\mathrm{new}} + \Theta_{t-1}\odot M_t^{\mathrm{old}},
  \label{eq:unbounded_full}
\end{equation}
where \(\odot\) denotes element-wise multiplication. The common ablations are
\begin{align}
  \Theta_t &= \Delta_t\odot M_t^{\mathrm{new}} + \Theta_{t-1}, &&\text{Only-New},\label{eq:onlynew}\\
  \Theta_t &= \Delta_t + \Theta_{t-1}\odot M_t^{\mathrm{old}}, &&\text{Only-Old}.\label{eq:onlyold_unbounded}
\end{align}
The previous self-modulation study found that old-state modulation is a key driver of the performance gain, because it directly controls how past fast-weight updates are retained, suppressed, amplified, or sign-reversed. This motivates bounding exactly the old-state branch rather than changing the full model.

\subsection{Bounded old-state modulation}
\label{subsec:bounded_old}

The raw entries of \(M_t^{\mathrm{old}}\) are unconstrained, because \cref{eq:gate} is an outer product of affine-head outputs. We therefore replace the recurrent old-state gate by
\begin{equation}
  \widetilde{M}_t^{\mathrm{old}}
  =
  \tanh\!\left(M_t^{\mathrm{old}}\right),
  \qquad
  |[\widetilde{M}_t^{\mathrm{old}}]_{kq}|\le 1
  \quad
  \forall k\in\{1,\ldots,L\},\ q\in\{1,\ldots,Q\},\ t.
  \label{eq:tanh_gate}
\end{equation}
The bounded full and bounded Only-Old recurrences are
\begin{align}
  \Theta_t &= \Delta_t\odot M_t^{\mathrm{new}}
            + \Theta_{t-1}\odot \widetilde{M}_t^{\mathrm{old}},
            &&\text{bounded full},\label{eq:bounded_full}\\
  \Theta_t &= \Delta_t
            + \Theta_{t-1}\odot \widetilde{M}_t^{\mathrm{old}},
            &&\text{bounded Only-Old}.\label{eq:bounded_onlyold}
\end{align}
No bound is applied to Standard QFWP or Only-New, because those variants do not multiply \(\Theta_{t-1}\) by an input-dependent gate. We also leave \(M_t^{\mathrm{new}}\) unchanged in the full model so that the modification is isolated to recurrent memory. The \(\tanh\) choice is sign-preserving, smooth, and close to the identity near zero; it bounds large recurrent multipliers without converting the old-state branch into a nonnegative sigmoid gate.

For a single coordinate \(j=(k,q)\), let
\(a_{t,j}=\widetilde{M}^{\mathrm{old}}_{t,j}\). The bounded Only-Old
recurrence unrolls as
\begin{equation}
  \theta_{t,j}=\theta_{0,j}\prod_{u=1}^{t}a_{u,j}
    + \sum_{s=1}^{t} d_{s,j}\prod_{u=s+1}^{t}a_{u,j},
  \qquad |a_{u,j}|\le 1,
  \label{eq:bounded_kernel}
\end{equation}
where \(d_{s,j}=[\Delta_s]_j\). Thus the recurrent kernel cannot geometrically amplify a stored update as any growth comes from the additive sequence of new updates rather than from repeated old-state multiplication.

\section{Quantum-Dynamics Prediction Tasks}
\label{sec:datasets}

We evaluate two quantum-dynamics benchmarks generated with CUDA-Q Dynamics, the dynamics-simulation backend of CUDA-Q~\cite{kim2023cuda}. In both cases, we extract a single scalar observable from a simulated quantum trajectory and treat it as a univariate next-step prediction task. Each trajectory contains 3000 evenly sampled time steps, is min--max normalized to \([-1,1]\), and is converted into chronological sliding-window samples: given \([x_{t-N},\ldots,x_{t-1}]\), the model predicts \(x_t\). The samples are split chronologically into 80\% training and 20\% test data.

\paragraph{Open Jaynes--Cummings dynamics.}
The first benchmark is an open Jaynes--Cummings system with a two-level qubit coupled to a single cavity mode truncated to five Fock levels. The Hamiltonian is
\begin{equation}
  H = \omega_c a^{\dagger}a + \omega_q \sigma_+\sigma_-
      + g(\sigma_-a^{\dagger}+\sigma_+a),
\end{equation}
with \(\omega_c=\omega_q=2\pi\) and \(g=\pi\). Photon loss is included through the collapse operator \(C=\sqrt{\gamma}\,a\), where \(\gamma=0.05\). The system is initialized as \(\rho_0=|g,1\rangle\langle g,1|\), and the prediction target is the qubit excitation probability \(\langle \sigma_+\sigma_-\rangle(t)\) over \(t\in[0,50]\).

\paragraph{Dispersive Transmon--resonator dynamics.}
The second benchmark is a closed dispersive transmon--resonator model, where the transmon is represented as a two-level system and the resonator is truncated to 20 Fock levels. The Hamiltonian is
\begin{equation}
  H=\frac{1}{2}\omega'_{01}\sigma_z+(\omega'_r+\chi\sigma_z)a^{\dagger}a .
\end{equation}
We use \(\omega_{01}=3.0\cdot 2\pi\) GHz, \(\omega_r=2.0\cdot 2\pi\) GHz, \(\chi=0.025\cdot 2\pi\) GHz, \(\omega'_{01}=\omega_{01}+\chi\), and \(\omega'_r=\omega_r\). The initial state is \((|0\rangle+|1\rangle)/\sqrt{2}\) for the transmon and \(|\alpha=2.0\rangle\) for the resonator. The prediction target is the resonator position quadrature \(\langle \hat{x}\rangle(t)\) over \(t\in[0,25]\) ns.

\section{Telecommunication Activity Prediction}
We also evaluate the Self-Modulating QFWP on a practical application scenario. Here we consider the Milan Telecommunication Activity Dataset~\cite{barlacchi2015multi}, a real-world urban spatiotemporal dataset collected in Milan, Italy. The dataset records telecommunication activity over spatial grid cells and includes multiple service modalities, such as SMS, call, and Internet traffic. In this work, we focus on the SMS activity signal and formulate the task as a univariate forecasting problem over individual spatial cells.

Each spatial cell is treated as a separate time series, and historical SMS activity values are used to predict future activity. In our experiments, we evaluate 100 spatial cells, corresponding to 100 SMS activity time series, for each model configuration. This setting provides a realistic benchmark for evaluating sequential forecasting models on urban telecommunication dynamics, where the temporal patterns may contain both short-term fluctuations and longer-range dependencies. Since our goal is to study the effect of self-modulation under different input window lengths, we restrict the experiments to the SMS modality and do not consider multimodal fusion in this paper. Although this benchmark is derived from a telecommunication forecasting task, the SMS activity sequences are real-world urban time series with noise, irregular fluctuations, and heterogeneous temporal patterns, making them useful for evaluating the potential applicability of self-modulating sequence models to other noisy real-world forecasting scenarios.

\section{Experimental Protocol}
\label{sec:setup}

We evaluate Standard QFWP, full Self-Modulating QFWP, Only-Old, and Only-New as inherited baselines, and we run the bounded-old modification for the multiplicative variants in the long-sequence stability study. The grid is
\begin{equation}
  H=Q\in\{4,6,8,10,12,14\},
  \qquad
  N\in\{4,8,16,32,64\},
\end{equation}
for 30 configurations per variant per dataset. All runs for quantum dynamics tasks use \(L=5\) variational layers, batch size 4, Adam optimizer~\cite{kingma2014adam} with learning rate \(10^{-3}\), gradient clipping with maximum \(\ell_2\)-norm 1.0, 100 training epochs, and seed 42. For telecommunication activity problem, we use \(L=2\) and batch size 16. A cell is treated as complete only when it reaches epoch 100 without a cancellation sidecar. Final test mean-squared error (MSE) is the primary accuracy metric; medians are emphasized where means are distorted by isolated divergent cells.

For variant comparisons, the relative improvement over Standard QFWP is
\begin{equation}
  \Delta_{\mathrm{rel}} = \frac{M_{\mathrm{standard}}-M_{\mathrm{variant}}}{M_{\mathrm{standard}}+\epsilon},
  \qquad \epsilon=10^{-12}.
\end{equation}
Positive values indicate lower MSE than the standard baseline. We also use Relative Strength, \(\Delta_{\mathrm{old}}-\Delta_{\mathrm{new}}\), to distinguish old-state and new-update modulation, and Synergy, \(\Delta_{\mathrm{both}}-\max(\Delta_{\mathrm{old}},\Delta_{\mathrm{new}})\), to test whether the full model exceeds the stronger single-sided variant. The bounded-vs-unbounded stability study is kept separate from the unbounded reproduction aggregates, because \cref{eq:bounded_full,eq:bounded_onlyold} define a modified architecture.

\section{Results}
\label{sec:results}

\subsection{Quantum dynamics prediction}
\label{subsec:prediction_quality}

\begin{figure}[tbp]
\centering
\includegraphics[width=\textwidth]{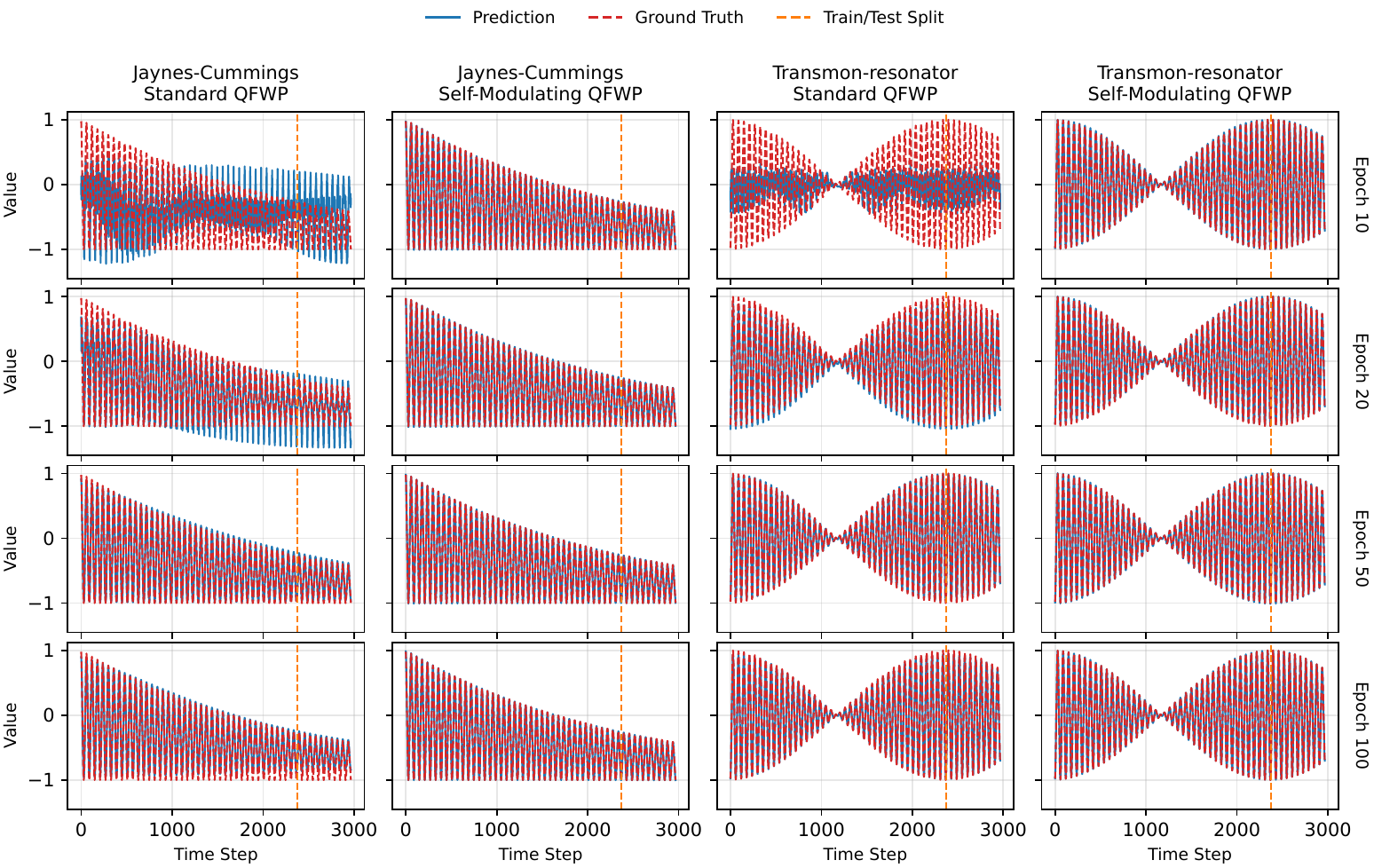}
\caption{Prediction trajectories at hidden size $H=10$ and sequence length $N=32$ on both benchmarks. Rows show epochs $10$, $20$, $50$, and $100$; columns compare Standard QFWP and full Self-Modulating QFWP on Jaynes--Cummings and transmon-resonator dynamics.}
\label{fig_fit}
\vspace{-15pt}
\end{figure}

During the unbounded sweep, gradient divergence was observed only in the multiplicative self-modulating variants, namely full Self-Modulating QFWP and Only-Old. These failures occur mainly at long sequence length and larger hidden size. We therefore analyze the completed unbounded cells as a reproduction of the prior self-modulating behavior, and separately evaluate the $\tanh$-bounded old-state gate on the long-sequence sub-grid $N\in{32,64}$.
The trajectory plots in \cref{fig_fit} show the practical effect of adaptive fast-weight memory on the two quantum-dynamics signals. On Jaynes--Cummings dynamics, Standard QFWP initially gives a noisy and amplitude-mismatched response, whereas the full Self-Modulating QFWP tracks the oscillatory decay much earlier in training. On the transmon-resonator signal, Standard QFWP shows early amplitude collapse, while the self-modulating model already follows the envelope and phase structure.

\begin{figure}[tbp]
\centering
\includegraphics[width=\textwidth]{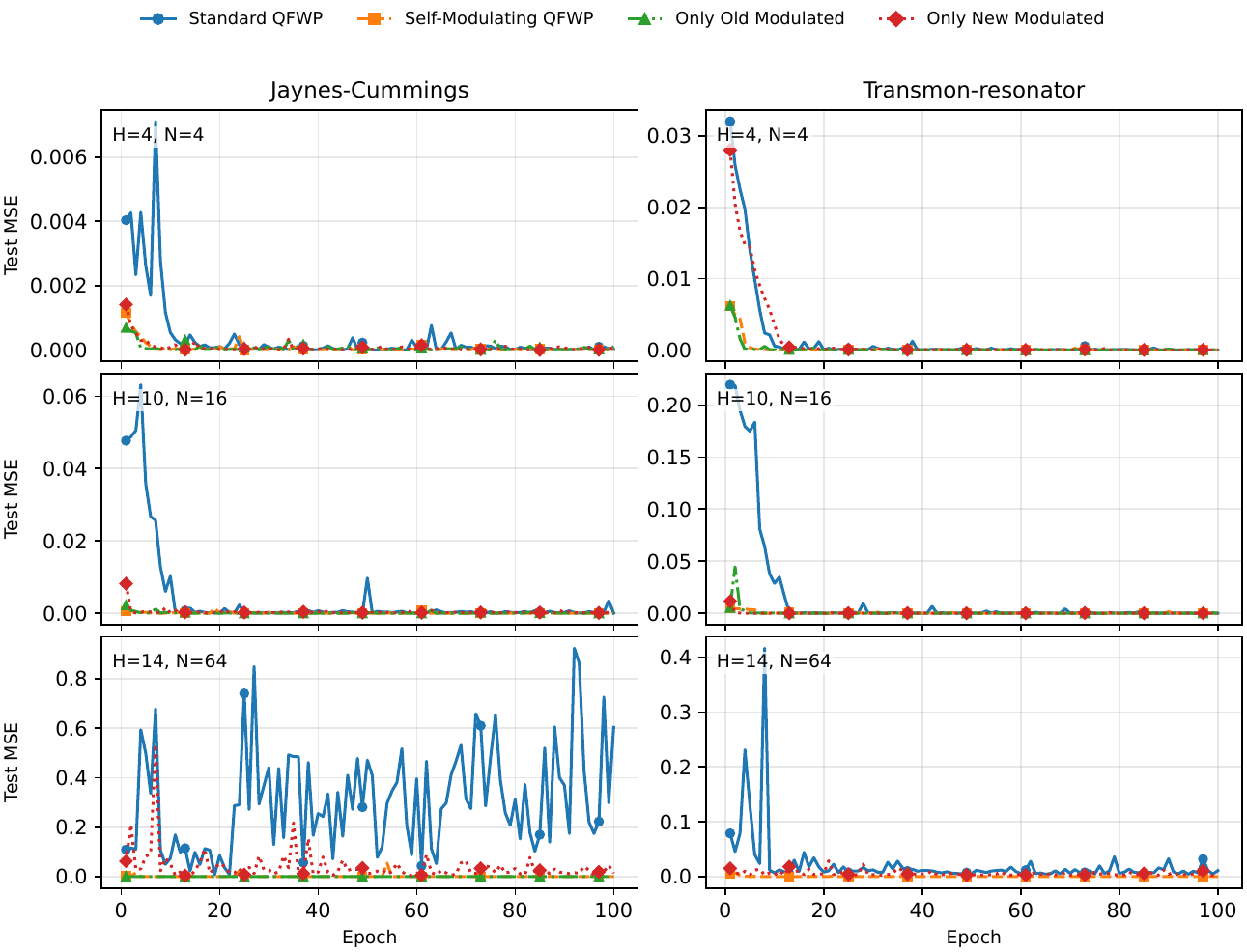}
\caption{Representative test-MSE convergence curves for $(H,N)=(4,4)$, $(10,16)$, and $(14,64)$. Columns show Jaynes--Cummings and transmon-resonator dynamics; curves compare Standard QFWP, full Self-Modulating QFWP, Only-Old, and Only-New.}
\label{fig_converge}
\vspace{-15pt}
\end{figure}

\begin{figure}[tbp]
\centering
\includegraphics[width=\textwidth]{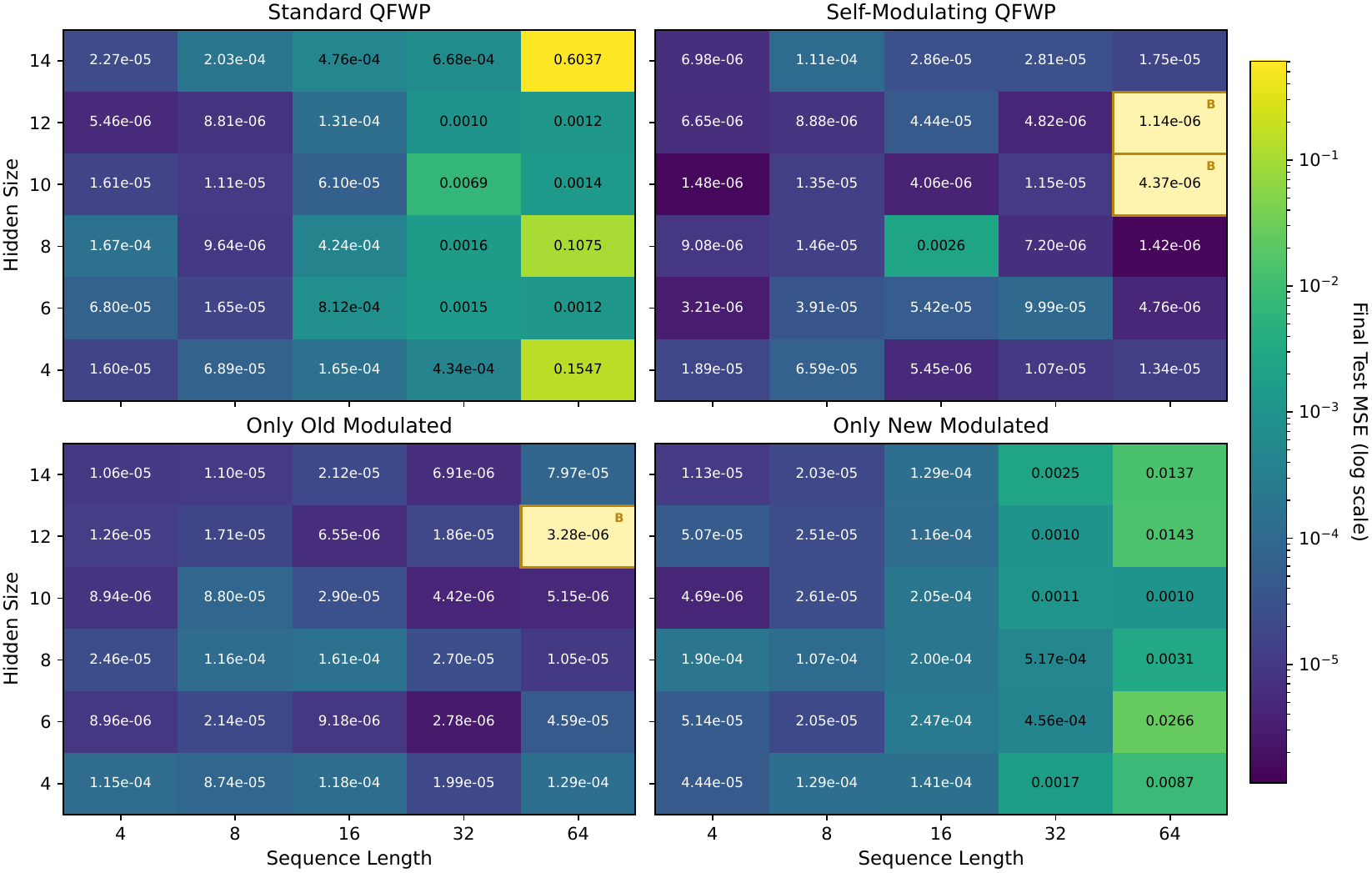}
\caption{Final test MSE on Jaynes--Cummings dynamics across hidden sizes and sequence lengths. Values are shown on a log scale. Cells marked \textbf{B} are unbounded multiplicative cells that did not reach epoch~$100$; their displayed values are replaced by the matched $\tanh$-bounded old-state run.}
\label{fig:jc_mse}
\vspace{-15pt}
\end{figure}

\begin{figure}[tbp]
\centering
\includegraphics[width=\textwidth]{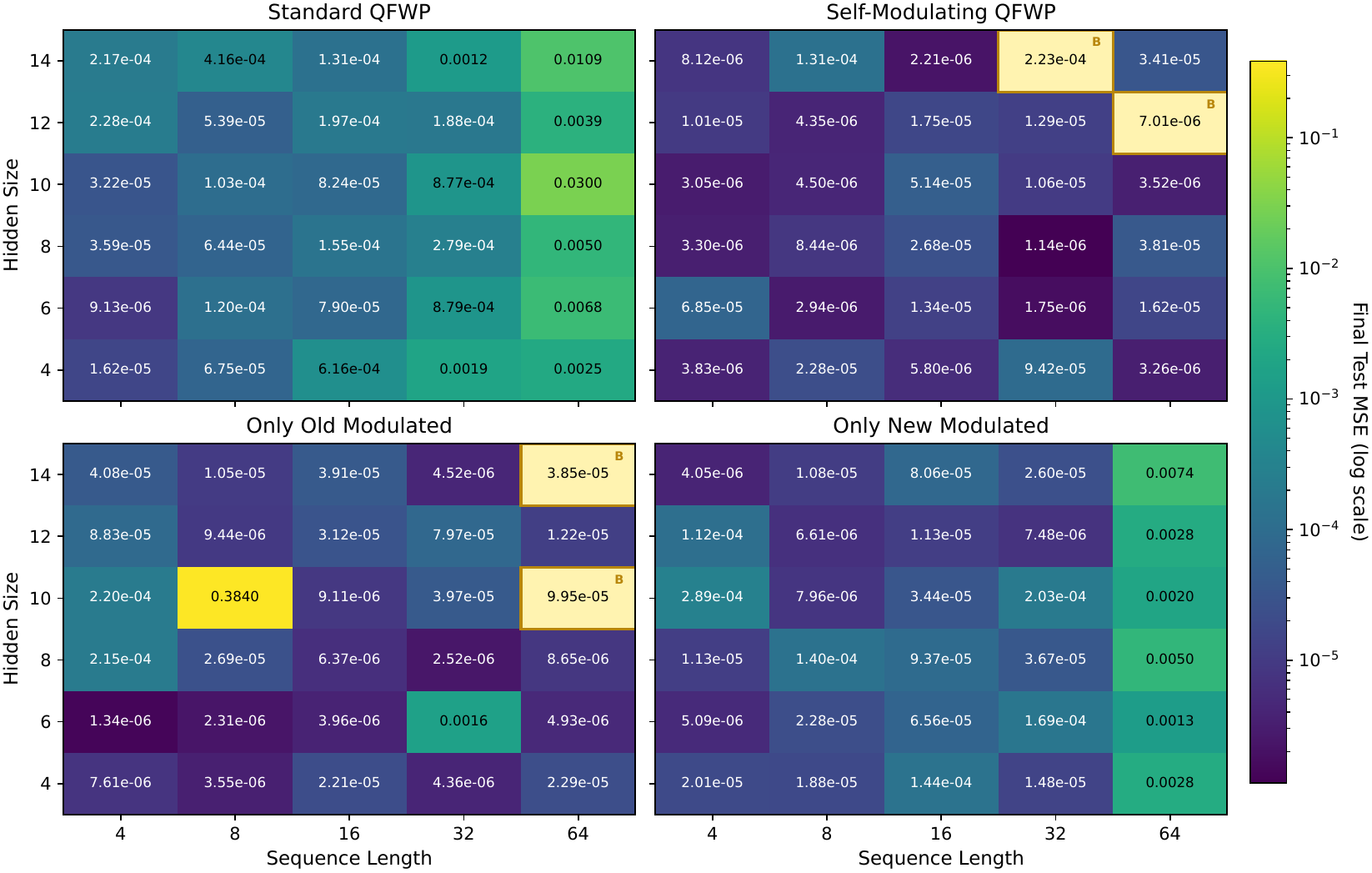}
\caption{Final test MSE on transmon-resonator dynamics across hidden sizes and sequence lengths. Values are shown on a log scale. Cells marked \textbf{B} are unbounded multiplicative cells replaced by the matched $\tanh$-bounded old-state run.}
\label{fig:transmon_mse}
\vspace{-15pt}
\end{figure}

The convergence curves in~\cref{fig_converge} are consistent with the trajectory plots. Across the completed unbounded cells, the modulated variants generally converge faster and reach lower test MSE than Standard QFWP, especially when the sequence is longer and the recurrent fast-weight state carries more history. This agrees with the main finding of the prior Self-Modulating QFWP study~\cite{chen2026selfmod}: modulation of the accumulated fast-weight state improves both optimization speed and predictive accuracy. The final-MSE heatmaps in \cref{fig:jc_mse,fig:transmon_mse} show that the main advantage of self-modulation appears most clearly at longer sequence lengths. In the rightmost columns, Standard QFWP often has substantially higher final MSE, indicating that the additive fast-weight update alone is less effective when the model must use a longer temporal context. Only-New modulation does not consistently resolve this degradation: although it improves some cells, it often remains close to the Standard QFWP error pattern at $N=32$ and $N=64$. In contrast, Only-Old and full Self-Modulating QFWP recover low-MSE solutions across most long-sequence configurations, showing that direct modulation of the accumulated fast-weight state is the dominant factor behind the improved prediction performance.

\begin{figure}[tbp]
\centering
\includegraphics[width=\textwidth]{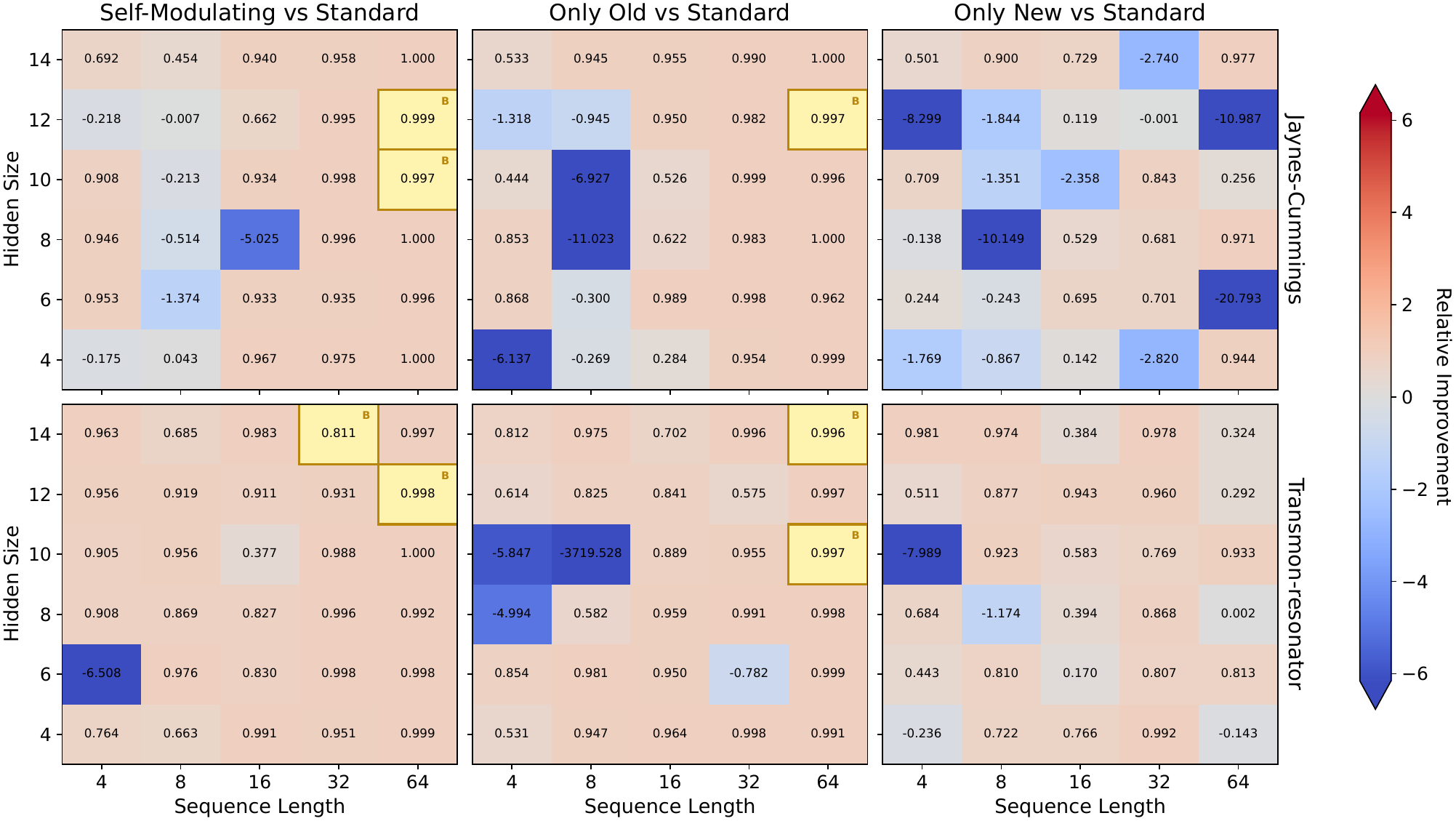}
\caption{Relative improvement over Standard QFWP for the three modulated variants. Positive values indicate lower MSE than Standard. Cells marked \textbf{B} use the matched bounded run because the corresponding unbounded multiplicative run did not complete; these cells are excluded from the completed-cell counts.}
\label{fig_improve}
\end{figure}

The same heatmaps also identify the stability limitation of the unbounded multiplicative gate. Divergence occurs only in the old-state multiplicative variants, namely full Self-Modulating QFWP and Only-Old. The affected cells are full Self-Modulating QFWP at $(H,N)=(12,64)$, $(10,64)$, Only-Old at $(12,64)$ in the Jaynes--Cummings benchmark, full Self-Modulating QFWP at $(14,32)$, $(12,64)$, and Only-Old at $(14,64)$, $(10,64)$ in the transmon-resonator benchmark. These cells are marked by \textbf{B}, where the displayed value is the matched bounded run rather than the failed unbounded run. The bounded replacements recover low final MSE in these otherwise unstable configurations, suggesting that the performance gain of old-state modulation can be retained while removing the long-sequence divergence mode. The relative-improvement heatmaps in \cref{fig_improve} separate the broad trend from isolated unstable cells. On Jaynes--Cummings dynamics, full Self-Modulating QFWP and Only-Old each improve over Standard QFWP in 21 of 28 all-four-completed cells, while Only-New improves in 15 of 28. On the transmon-resonator benchmark, full Self-Modulating QFWP improves in 25 of 26 all-four-completed cells, and Only-Old and Only-New each improve in 22 of 26. A few negative ratios arise either from unstable cells or from very small absolute differences when the Standard baseline already has near-zero error. We therefore treat the ratio heatmaps as diagnostics to be read together with the raw MSE.

\begin{figure}[tbp]
\centering
\includegraphics[width=.98\textwidth]{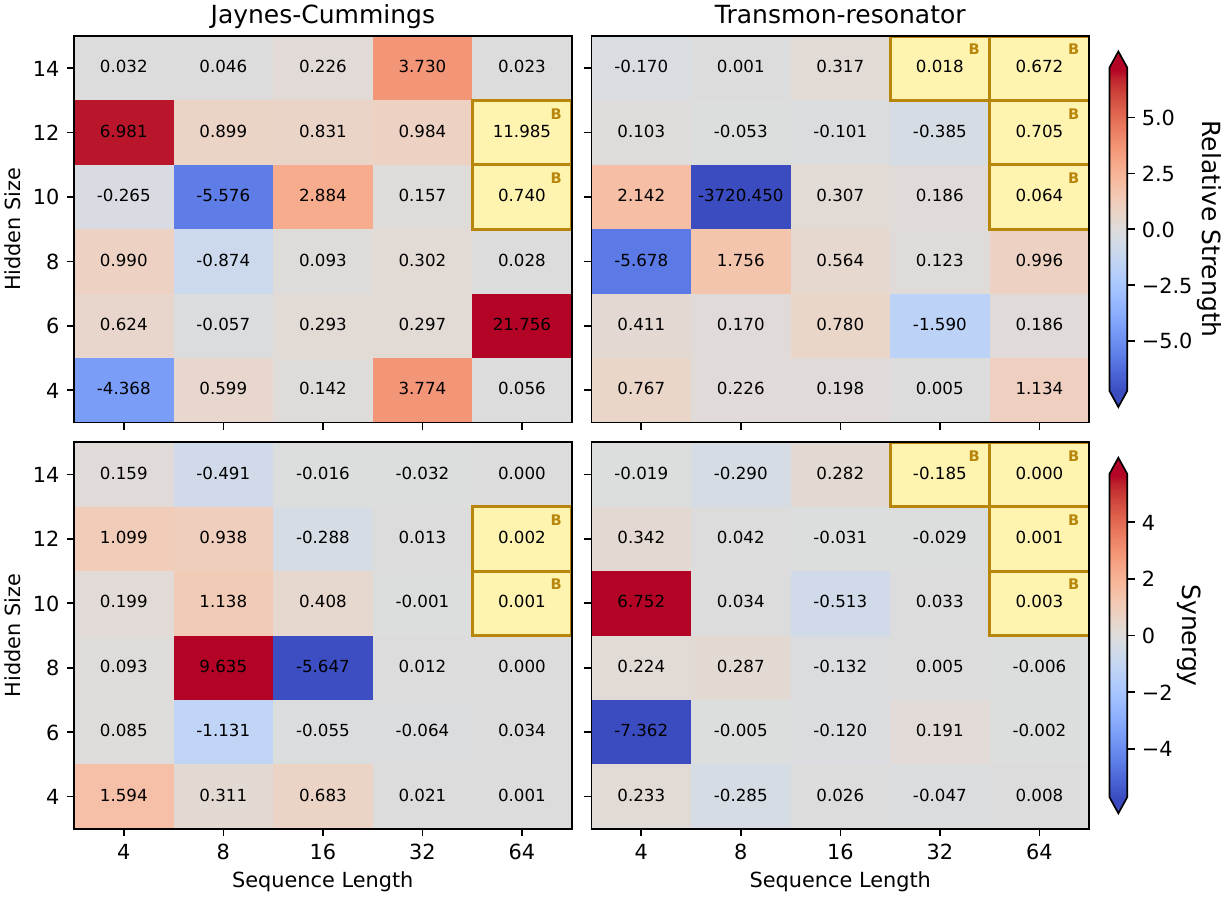}
\caption{Old-state dominance and synergy diagnostics. Positive Relative Strength means that old-state modulation improves more than new-update modulation; positive Synergy means that the full model exceeds the better single-sided variant.}
\label{fig_syn}
\vspace{-10pt}
\end{figure}
The Relative Strength and Synergy diagnostics in \cref{fig_syn} provide a qualitative view of how the two modulation branches contribute. The Relative Strength panels show that many configurations favor old-state modulation over new-update modulation, especially in cells where the long-sequence degradation of Standard QFWP is recovered. This is consistent with the heatmap evidence that direct control of the accumulated fast-weight state is often the more important component. The Synergy panels, however, are more mixed: the full model does not uniformly exceed the better single-sided variant. We therefore interpret the full Self-Modulating QFWP primarily as a robust combination of old-state and new-update gates, while the most consistent performance gain comes from input-dependent modulation of the accumulated fast-weight state.

\begin{figure}[tbp]
\centering
\includegraphics[width=.98\textwidth]{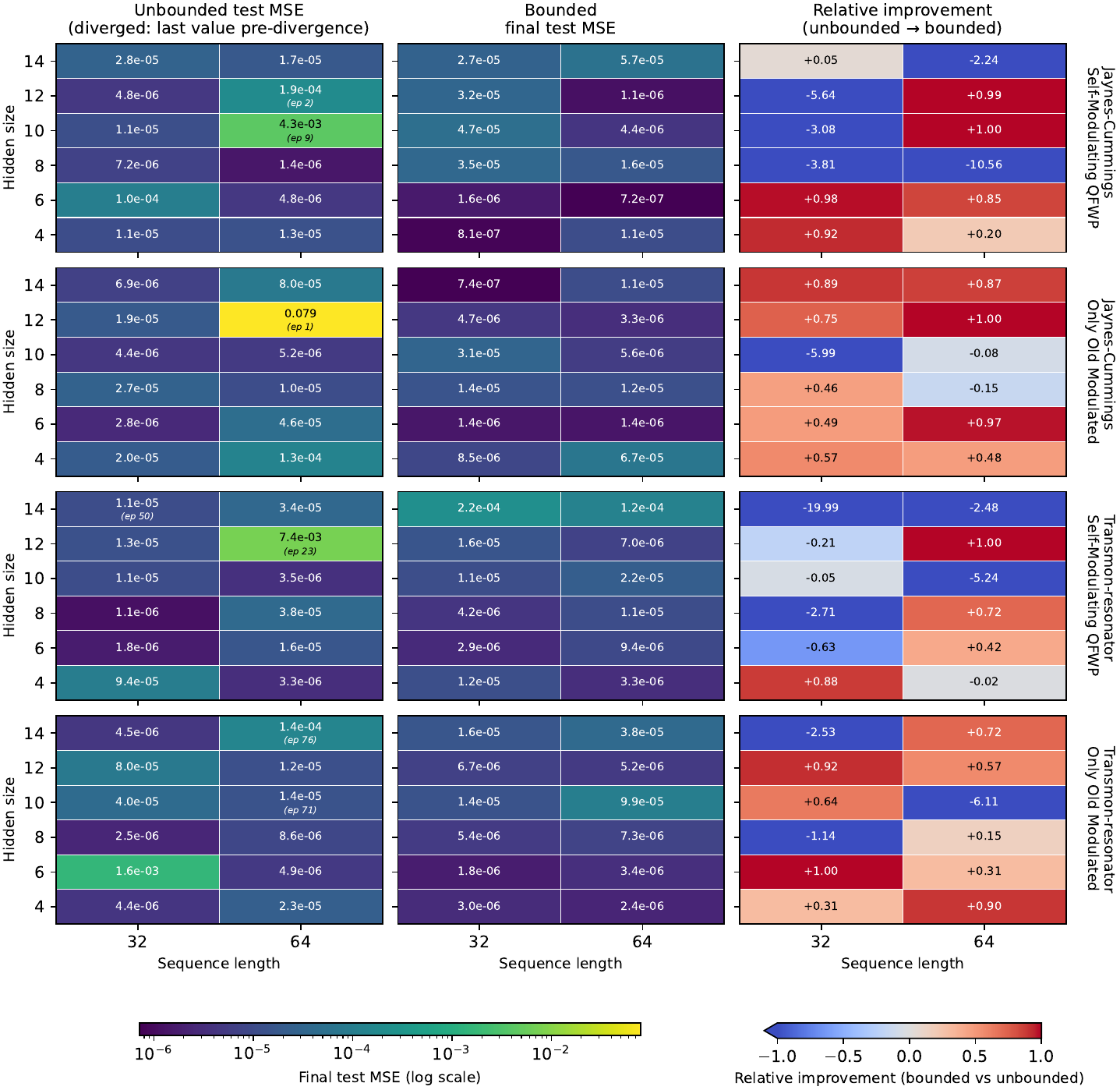}
\caption[Bounded versus unbounded multiplicative variants on the long-sequence sub-grid.]{
Bounded versus unbounded multiplicative variants on the long-sequence sub-grid. The left column shows the unbounded test MSE, and the middle column shows the matched \(\tanh\)-bounded final test MSE. Annotations of the form \((\mathrm{ep};k)\) mark unbounded cells that diverged at epoch \(k\); for these cells, the displayed value is the effective test MSE from the last finite epoch before divergence.

}

\label{fig_bound}
\vspace{-15pt}
\end{figure}

\begin{table}[tbp]
  \centering
  \caption{Long-sequence bounded-vs-unbounded summary over $N\in{32,64}$. Means and medians are computed over 12 cells per dataset and variant. Bounded columns apply only to the multiplicative variants; divergent unbounded cells use the last finite test MSE before termination, with the source epoch annotated in \cref{fig_bound}.}
  \label{tab:bound}
  \begin{tabular}{llcccc}
    \toprule
    & & \multicolumn{2}{c}{Unbounded} & \multicolumn{2}{c}{Bounded} \\
    \cmidrule(lr){3-4}\cmidrule(lr){5-6}
    Dataset & Variant & Mean & Median & Mean & Median \\
    \midrule
    Jaynes--Cummings & Standard QFWP & \(7.35\times10^{-2}\) & \(1.46\times10^{-3}\) & --- & --- \\
     & Self-Modulating & \(3.91\times10^{-4}\) & \(1.24\times10^{-5}\) & \(1.94\times10^{-5}\) & \(1.35\times10^{-5}\) \\
     & Only-Old & \(6.63\times10^{-3}\) & \(1.93\times10^{-5}\) & \(1.34\times10^{-5}\) & \(7.02\times10^{-6}\) \\
     & Only-New & \(6.22\times10^{-3}\) & \(2.08\times10^{-3}\) & --- & --- \\
    \midrule
    Transmon-resonator & Standard QFWP & \(5.37\times10^{-3}\) & \(2.18\times10^{-3}\) & --- & --- \\
     & Self-Modulating & \(6.37\times10^{-4}\) & \(1.17\times10^{-5}\) & \(3.66\times10^{-5}\) & \(1.08\times10^{-5}\) \\
     & Only-Old & \(1.58\times10^{-4}\) & \(1.31\times10^{-5}\) & \(1.69\times10^{-5}\) & \(6.07\times10^{-6}\) \\
     & Only-New & \(1.81\times10^{-3}\) & \(7.37\times10^{-4}\) & --- & --- \\
    \bottomrule
  \end{tabular}
\end{table}

The bounded-vs-unbounded comparison in \cref{tab:bound,fig_bound} shows that
the numerical instability is tied to the recurrent old-state multiplication.
The \(\tanh\)-bounded gate completes all long-sequence cells and reduces the
mean final MSE for both old-state multiplicative variants on both datasets.
It also improves the median in three of the four dataset--variant pairs; the
only exception is the Jaynes--Cummings full Self-Modulating row, where the
median changes only slightly from \(1.24\times10^{-5}\) to
\(1.35\times10^{-5}\).

The per-cell comparison should be interpreted as a stability diagnostic rather
than as a claim that the bounded run is lower in every cell.  The largest
positive gains occur in failed high-error unbounded cells such as for the Jaynes--Cummings dataset, full Self-Modulating QFWP at \((H,N)=(12,64)\) and
\((10,64)\), Only-Old at \((12,64)\), and for the transmon-resonator dataset, full
Self-Modulating QFWP at \((12,64)\), and Only-Old at \((14,64)\).
However, in the full Self-Modulating QFWP at \((14,32)\) and
Only-Old at \((10,64)\) for the transmon-resonator dataset, the bounded run completes with low absolute MSE but
is higher than the last finite pre-divergence unbounded value.  Thus, the
main conclusion is that bounding \(M_t^{\mathrm{old}}\) removes the
long-sequence failure mode and improves aggregate robustness, while preserving
the input-dependent temporal filtering mechanism in
\cref{eq:bounded_kernel}.

\subsection{Telecommunication Activity Prediction}
\begin{figure}[tbp]
\centering
\includegraphics[width=1\textwidth]{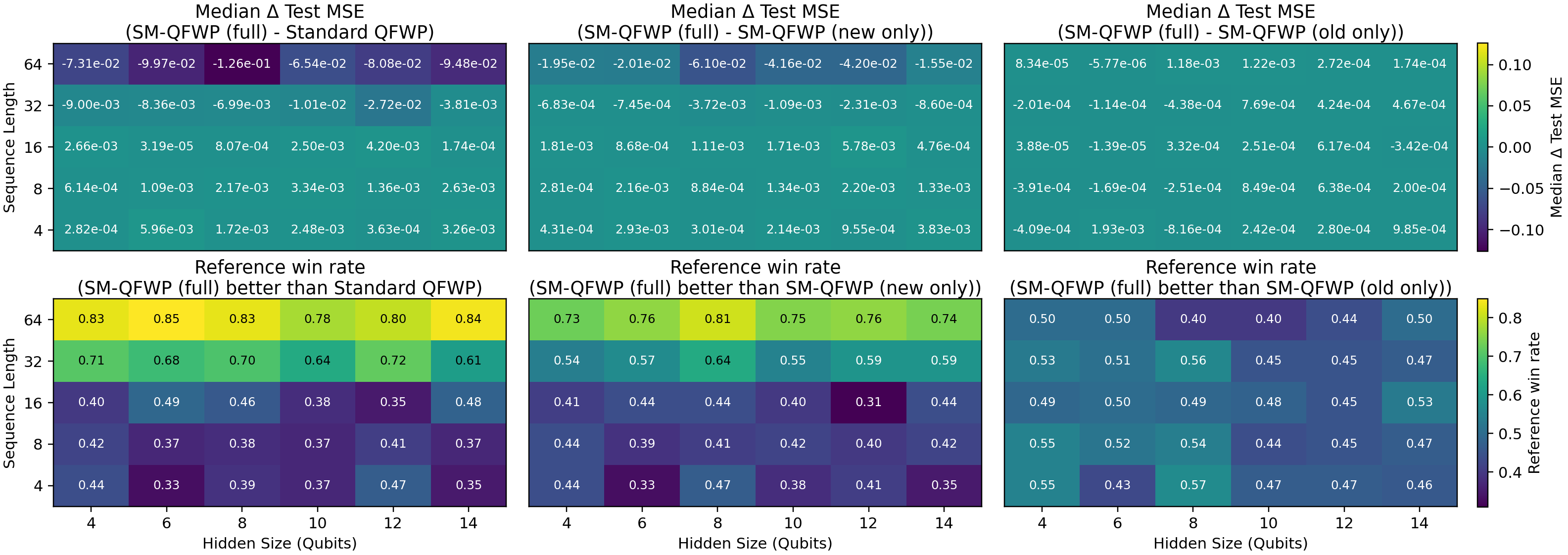}
\caption[Paired Test MSE comparison of full SM-QFWP against Standard QFWP and two self-modulating ablations on Milan SMS forecasting.]{
Each cell reports results for a fixed sequence length and hidden size. The top row shows the median paired difference in Test MSE, computed as full SM-QFWP minus the comparison method, where negative values indicate better performance of full SM-QFWP. The bottom row shows the paired win rate of full SM-QFWP across square IDs. Full SM-QFWP exhibits clear gains over Standard QFWP and the new-parameter-only ablation primarily at longer sequence lengths, especially for sequence length 64, while its performance is close to the old-parameter-only ablation.
}

\label{fig_sms_telecom}
\vspace{-15pt}
\end{figure}

In this part of the experiment, we evaluate the Self-Modulating QFWP on Milan Telecommunication Activity Dataset. Each cell represents the aggregated results from 100 time-series training, there 100 individually trained models. The paired comparison shown in \cref{fig_sms_telecom} reveals that the benefit of self-modulation is strongly sequence-length dependent. For short input windows, full Self-Modulating QFWP does not consistently outperform the baselines; however, as the sequence length increases, especially at sequence length 64, full Self-Modulating QFWP achieves consistently lower Test MSE than both Standard QFWP and the Only-New ablation, with paired win rates mostly above 0.7. This suggests that self-modulation becomes most useful when the forecasting task requires longer temporal context.

Notably, the full model performs similarly to the Only-Old ablation, while the Only-New ablation is clearly weaker in the long-sequence regime. This ablation result is consistent with our previous observations: the main advantage of self-modulation comes from modulating the memory-bearing existing QFWP parameters, rather than merely adding a new modulated parameter branch. These numerical results therefore support the view that retaining and modulating internal memory is critical for exploiting long-range temporal structure in the Milan SMS forecasting task. 

Finally, we note that the Milan SMS experiment uses the original unbounded Self-Modulating QFWP and its ablations, rather than the bounded-old modification. This choice reflects the role of this benchmark in our study. The SMS task contains a much larger number of independently trained time series, and all tested configurations completed training without requiring the bounded old-state gate. Therefore, we use this benchmark primarily to test whether the self-modulating memory mechanism observed in the quantum-dynamics experiments also transfers to noisy real-world urban time series. The results support this interpretation: the gains of full Self-Modulating QFWP become most
consistent at longer input windows, and its behavior remains close to the Only-Old ablation, whereas Only-New is weaker in the long-sequence regime. Thus, the SMS results provide practical evidence that modulating the accumulated fast-weight state is useful beyond simulated quantum dynamics. At the same time, because bounded variants were not exhaustively evaluated on this larger telecommunication grid, we do not claim that the bounded gate is unnecessary for all real-world forecasting settings. Rather, the bounded operation should be viewed as a stability mechanism for regimes where unbounded old-state multiplication exhibits divergence, while the SMS benchmark demonstrates that the original self-modulating mechanism can already be effective and stable under the present telecommunication forecasting protocol.
\section{Conclusion}
\label{sec:conclusion}

We studied Self-Modulating Quantum Fast Weight Programmers as adaptive
fast-weight memory controllers. Experiments on two CUDA-Q quantum-dynamics forecasting benchmarks show that the main benefit of self-modulation comes from input-dependent control of the accumulated fast-weight state, with Only-Old and full Self-Modulating QFWP providing the most consistent long-sequence improvements over Standard QFWP. At the same time, the unbounded old-state multiplier can diverge in difficult long-sequence regimes. The proposed sign-preserving \(\tanh\)-bounded old-state gate removes this failure mode by bounding only the recurrent memory branch while leaving the new-update branch unchanged. The Milan SMS telecommunication results further show that the original Self-Modulating QFWP transfers to noisy real-world time series and is most useful at longer input windows, where its behavior remains close to the Only-Old ablation. Together, these findings identify accumulated-memory modulation as the key mechanism of Self-Modulating QFWP and bounded old-state gating as a simple stabilization strategy when unbounded recurrence becomes
unstable.

\begin{credits}
\subsubsection{\ackname} The views expressed in this article are those of the authors and do not represent the views of Wells Fargo. This article is for informational purposes only. Nothing contained in this article should be construed as investment advice. Wells Fargo makes no express or implied warranties and expressly disclaims all legal, tax, and accounting implications related to this article.

\end{credits}
\bibliographystyle{splncs04}
\bibliography{bib/ref,bib/fwp,bib/qlstm,bib/qml}

\begin{thebibliography}{10}
\providecommand{\url}[1]{\texttt{#1}}
\providecommand{\urlprefix}{URL }
\providecommand{\doi}[1]{https://doi.org/#1}

\bibitem{anschuetz2023interpretable}
Anschuetz, E.R., Hu, H.Y., Huang, J.L., Gao, X.: Interpretable quantum
  advantage in neural sequence learning. PRX Quantum  \textbf{4}(2),  020338
  (2023)

\bibitem{barlacchi2015multi}
Barlacchi, G., et~al.: A multi-source dataset of urban life in the city of
  milan and the province of trentino. Scientific data  \textbf{2}(1),  1--15
  (2015)

\bibitem{bausch2020recurrent}
Bausch, J.: Recurrent quantum neural networks. Advances in neural information
  processing systems  \textbf{33},  1368--1379 (2020)

\bibitem{cao2023linear}
Cao, Y., Zhou, X., Fei, X., Zhao, H., Liu, W., Zhao, J.: Linear-layer-enhanced
  quantum long short-term memory for carbon price forecasting. Quantum Machine
  Intelligence  \textbf{5}(2), ~26 (2023)

\bibitem{ceschini2026qfwp}
Ceschini, A., Rosato, A., Panella, M., Chen, S.Y.C.: Quantum fast weight
  programming for time series prediction. In: ICASSP 2026-2026 IEEE
  International Conference on Acoustics, Speech and Signal Processing (ICASSP).
  pp. 22032--22036. IEEE (2026)

\bibitem{chen2025benchmarking}
Chen, C.S., Chen, S.Y.C., Tsai, Y.C.: Benchmarking quantum and classical
  sequential models for urban telecommunication forecasting. arXiv preprint
  arXiv:2508.04488  (2025)

\bibitem{chen2025toward}
Chen, K.C., Chen, S.Y.C., Liu, C.Y., Leung, K.K.: Toward large-scale
  distributed quantum long short-term memory with modular quantum computers.
  In: 2025 International Wireless Communications and Mobile Computing (IWCMC).
  pp. 337--342. IEEE (2025)

\bibitem{chen2024qfwp}
Chen, S.Y.C.: Learning to program variational quantum circuits with fast
  weights. In: 2024 International Joint Conference on Neural Networks (IJCNN).
  pp.~1--9. IEEE (2024)

\bibitem{chen2022qlstm}
Chen, S.Y.C., Yoo, S., Fang, Y.L.L.: Quantum long short-term memory. In: Icassp
  2022-2022 IEEE international conference on acoustics, speech and signal
  processing (ICASSP). pp. 8622--8626. IEEE (2022)

\bibitem{chen2026recursive_qlstm}
Chen, S.Y.C., et~al.: Recursive qlstm with dynamic variational quantum circuit
  adaptation (2026), \url{https://arxiv.org/abs/2606.24932}

\bibitem{chen2026selfmod}
Chen, S.Y.C., et~al.: Self-modulating quantum fast-weight programmers for
  efficient adaptive sequential learning (2026),
  \url{https://arxiv.org/abs/2606.24933}

\bibitem{hsu2025quantum}
Hsu, Y.C., Chen, N.Y., Li, T.Y., Lee, P.H.H., Chen, K.C.: Quantum kernel-based
  long short-term memory for climate time-series forecasting. In: 2025
  International Conference on Quantum Communications, Networking, and Computing
  (QCNC). pp. 421--426. IEEE (2025)

\bibitem{hsu2025federated_qkernel_lstm}
Hsu, Y.C., et~al.: Federated quantum kernel-based long short-term memory for
  human activity recognition. In: 2025 IEEE International Conference on Quantum
  Computing and Engineering (QCE). vol.~02, pp. 54--58 (2025).
  \doi{10.1109/QCE65121.2025.10293}

\bibitem{hsu2026qkanlstm}
Hsu, Y.C., et~al.: {QKAN-LSTM}: Quantum-inspired {Kolmogorov}--{Arnold} long
  short-term memory. In: 2026 International Conference on Quantum
  Communications, Networking, and Computing (QCNC). pp. 650--659. IEEE (2026)

\bibitem{irie2021going}
Irie, K., Schlag, I., Csord{\'a}s, R., Schmidhuber, J.: Going beyond linear
  transformers with recurrent fast weight programmers. Advances in neural
  information processing systems  \textbf{34},  7703--7717 (2021)

\bibitem{jiang2025qvaf_qkan}
Jiang, J.C., Huang, M.Y.C., Chen, T., Goan, H.S.: Quantum variational
  activation functions empower {Kolmogorov}-{Arnold} networks. arXiv preprint
  arXiv:2509.14026  (2025). \doi{10.48550/arXiv.2509.14026},
  \url{https://arxiv.org/abs/2509.14026}

\bibitem{khan2024solarqlstm}
Khan, S.Z., et~al.: Quantum long short-term memory (qlstm) vs. classical lstm
  in time series forecasting: a comparative study in solar power forecasting.
  Frontiers in Physics  \textbf{12},  1439180 (2024)

\bibitem{kim2023cuda}
Kim, J.S., et~al.: Cuda quantum: The platform for integrated quantum-classical
  computing. In: 2023 60th ACM/IEEE Design Automation Conference (DAC).
  pp.~1--4. IEEE (2023)

\bibitem{kingma2014adam}
Kingma, D.P., Ba, J.: Adam: A method for stochastic optimization. arXiv
  preprint arXiv:1412.6980  (2014)

\bibitem{li2024airqlstm}
Li, F., Dong, Y.: Air quality prediction based on improved quantum long
  short-term memory neural networks. Physica Scripta  \textbf{99}(8),  085035
  (2024)

\bibitem{lin2024qtlstm}
Lin, C.H.A., Liu, C.Y., Chen, K.C.: Quantum-train long short-term memory:
  Application on flood prediction problem. In: 2024 IEEE International
  Conference on Quantum Computing and Engineering (QCE). vol.~2, pp. 268--273.
  IEEE (2024)

\bibitem{lin2026gqkae}
Lin, Y.C., et~al.: Generative quantum-inspired {Kolmogorov}-{Arnold}
  eigensolver (2026), \url{https://arxiv.org/abs/2605.04604}

\bibitem{liu2025qtqfwp}
Liu, C.Y., Chen, S.Y.C., Chen, K.C., Huang, W.J., Chang, Y.J.: Programming
  variational quantum circuits with quantum-train agent. In: 2025 International
  Conference on Quantum Communications, Networking, and Computing (QCNC). pp.
  544--548. IEEE (2025)

\bibitem{peng2026qkanfwp}
Peng, K.C., et~al.: Gated qkan-fwp: Scalable quantum-inspired sequence
  learning. arXiv preprint arXiv:2605.06734  (2026)

\bibitem{peng2026qkanfwp_TM}
Peng, K.C., et~al.: Parameter-efficient quantum-inspired fast weight
  programmers for traffic-matrix forecasting (2026)

\bibitem{schlag2021linear}
Schlag, I., Irie, K., Schmidhuber, J.: Linear transformers are secretly fast
  weight programmers. In: International conference on machine learning. pp.
  9355--9366. PMLR (2021)

\bibitem{schmidhuber1992fast}
Schmidhuber, J.: Learning to control fast-weight memories: An alternative to
  dynamic recurrent networks. Neural Computation  \textbf{4}(1),  131--139
  (1992)

\bibitem{su2025blsqlstm}
Su, L., Li, D., Qiu, D.: Bls-qlstm: a novel hybrid quantum neural network for
  stock index forecasting. Humanities and Social Sciences Communications
  \textbf{12}(1),  1--15 (2025)

\bibitem{tran2025eqlstml}
Tran, B.N.D., et~al.: Quantum lstm model for estimation of energy expenditure
  in human aging using wearable iot healthcare technology. IEEE Internet of
  Things Journal  (2025)

\bibitem{tsurkan2025hqrnn}
Tsurkan, O., et~al.: Hybrid quantum recurrent neural network for remaining
  useful life prediction. arXiv preprint arXiv:2504.20823  (2025)

\bibitem{zhang2026quantum}
Zhang, L., Xu, Y., Wu, M., Wang, L., Xu, H.: Quantum long short-term memory for
  drug discovery. EPJ Quantum Technology  \textbf{13}(1), ~14 (2026)

\end{thebibliography}

\end{document}